\documentclass[secnumarabic,amssymb, nobibnotes, aps, prl,preprint]{revtex4-1}
\usepackage{graphicx}	
\usepackage{graphics}	
\usepackage{amsmath}
\usepackage{upgreek}
\usepackage[english]{babel}

\begin{document}
\title{A Coherent Polariton Laser}

\author{Seonghoon Kim$^1$, Bo Zhang$^2$, Zhaorong Wang$^1$, Julian Fischer$^3$, Sebastian Brodbeck$^3$, Martin Kamp$^3$, Christian Schneider$^3$, Sven H\"ofling$^{3,4}$, Hui Deng$^{2}$}
\small{ }
\address{$^1$ EECS Department, University of Michigan, 1301 Beal Avenue, Ann Arbor, MI 48109-2122, USA}
\address{$^2$ Physics Department, University of Michigan, 450 Church Street, Ann Arbor, MI 48109-1040, USA}
\address{$^3$ Technische Physik, Universit\"at W\"urzburg, Am Hubland, D-97074 W\"urzburg, Germany}
\address{$^{4}$SUPA, School of Physics and Astronomy, University of St Andrews, St Andrews KY16 9SS, United Kingdom}

\maketitle

\textbf{The semiconductor polariton laser promises a new source of coherent light, which, compared to conventional semiconductor photon lasers, has input-energy threshold orders of magnitude lower. However, intensity stability, a defining feature of a coherent state, has remained poor. Intensity noise at many times of the shot-noise of a coherent state has persisted, which has been attributed to multiple mechanisms that are difficult to separate in conventional polariton systems. The large intensity noise in turn limited the phase coherence. These limit the capability of the polariton laser as a source of coherence light. Here, we demonstrate a polariton laser with shot-noise limited intensity stability, as expected of a fully coherent state. This is achieved by using an optical cavity with high mode selectivity to enforce single-mode lasing, suppress condensate depletion, and establish gain saturation. The absence of spurious intensity fluctuations moreover enabled measurement of a transition from exponential to Gaussian decay of the phase coherence of the polariton laser. It suggests large self-interaction energies in the polariton condensate, exceeding the laser bandwidth. Such strong interactions are unique to matter-wave laser and important for nonlinear polariton devices. The results will guide future development of polariton lasers and nonlinear polariton devices.}

\section{\textsc{Introduction}}
Coherent light, characterized by long phase coherence and shot-noise limited intensity stability \cite{glauber_quantum_1963}, has been a revolutionary resource in modern science and technology. The coherence in conventional photon lasers is formed through the stimulated photon emission process that requires population inversion in the incoherent gain medium, which sets a lower bound for the energy threshold. 
In contrast, a coherent matter wave can form as the ground state, rather than inverted state, of a manybody quantum system. Therefore, semiconductor polaritons, half-light and half-matter quasi-particles that can form a condensate \cite{deng_exciton-polariton_2010}, promise an energy-efficient source of coherent light without electronic population inversion \cite{imamoglu_nonequilibrium_1996, deng_polariton_2003}, called the polariton laser.

Polariton condensation and lasing at a low threshold density has been demonstrated worldwide in diverse materials \cite{deng_polariton_2003,christopoulos_room-temperature_2007,bajoni_polariton_2008,kena-cohen_room-temperature_2010,lu_room_2012, Xie_room-temperature_2012, li_excitonic_2013} under both optical and electrical excitations \cite{schneider_electrically_2013,bhattacharya_solid_2013,bhattacharya_room_2014}. Evidence of polariton lasing in the polariton ground state includes quantum degeneracy \cite{deng_condensation_2002}, reduced intensity fluctuations compared to thermal states \cite{deng_condensation_2002}, increased spatial \cite{kasprzak_boseeinstein_2006} and temporal \cite{love_intrinsic_2008} coherence, and spontaneous polarization build-up \cite{baumberg_spontaneous_2008}.

In these polariton lasers, however, a significant fraction of thermal population were either clearly present in the ground state or could not be excluded. Consequently, phase coherence and intensity stability, the defining features of a coherent state that are crucial for many applications of lasers, have been limited.  
%
%
The intensity noise is measured by the second-order coherence function $g^{(2)}(0)$:
\begin{align}
  g^{(2)}(0) = 1 +\frac{\sigma_n^2 - \bar{n}}{\bar{n}^2},
  \label{eq:g2}
\end{align}
where $\sigma_n^2$ is the variance of the photon number and $\bar{n}$ is the average photon number. For coherent state with Poisson number statistics, $\sigma_n^2 =\bar{n}$ and $g^{(2)}(0)=1$, as has been observed in both photon lasers and coherent matter waves of atoms \cite{schellekens_hanbury_2005,ottl_correlations_2005,dall_observation_2011}.
In polariton lasers, however, large intensity noise has persisted, as measured by $g^{(2)}(0)>1$ \cite{deng_condensation_2002,kasprzak_second-order_2008, love_intrinsic_2008, horikiri_higher_2010, asmann_polariton_2011, tempel_characterization_2012, amthor_electro-optical_2014}. One of the well-calibrated values reached as low as $g^{(2)}(0)=1.1$, which still corresponded to a variance $\sigma_n^2$ that was 50 times of the shot-noise limit \cite{love_intrinsic_2008}. Such large intensity noise in polariton lasers has been attributed to multiple possible mechanisms, including mode competition among multiple spatial modes or momentum states \cite{kusudo_stochastic_2013}, reservoir-induced intensity fluctuations \cite{wouters_stochastic_2009}, and condensate depletion \cite{haug_temporal_2012, schwendimann_stationary_2010}. These different mechanisms coexisted in previous experiments and could not be separately identified or controlled. Agreements between experiments and theory were qualitative. No solution has been clearly identified to improve the intensity stability of a polariton laser. The quantum statistical nature of polariton lasers and their potential use as a source of coherent light were unclear.


The spurious intensity fluctuations in turn limits the phase coherence.
The phase coherence is described by the first-order coherence function $g^{(1)}(\tau)$, which equals 1 at $\tau=0$ and decays with a coherence time corresponding to the inverse linewidth of the light. In conventional photon laser, $g^{(1)}(\tau)$ decays exponentially and the coherence time increases proportionally with the photon occupation number, giving the Schawlow-Townes linewidth \cite{Schawlow_infrared_1958}. For polariton lasers, however, the coherence time was limited by energy fluctuations induced by pulsed excitation in early experiments. When intensity stabilized excitation sources were used, energy fluctuations resulting from the spurious intensity fluctuations of the condensate became a main dephasing mechanism \cite{love_intrinsic_2008}. Consequently, the intrinsic limit of phase coherence of the polariton laser remained obscured.

In this work, we demonstrate a coherent polariton laser with intensity and phase noise limited only by the intrinsic shot-noise of the condensate. We achieve intensity stability at the Poisson limit in a single-mode polariton laser with suppressed mode-competition and condensate depletion by a cavity with high mode-selectivity. Gain saturation was established to avoid intensity fluctuations induced by the reservoir and non-lasing modes. 
With full intensity stability, the intrinsic phase coherence of the polariton laser showed a transition from the Schawlow-Townes limit at low condensate occupancies to Gaussian-dephasing at high occupancies. Such a transition was predicted for matter-wave lasers but not observed in experiments before. The Gaussian dephasing results from strong interactions within the condensate, which is unique to a matter-wave lasers \cite{wiseman_defining_1997,thomsen_atom-laser_2002} and important for nonlinear polariton devices \cite{verger_polariton_2006,liew_optical_2008,liew_single_2010}.

\section{\textsc{The Cavity system}}

\begin{figure*}[tb]
\centering
\includegraphics{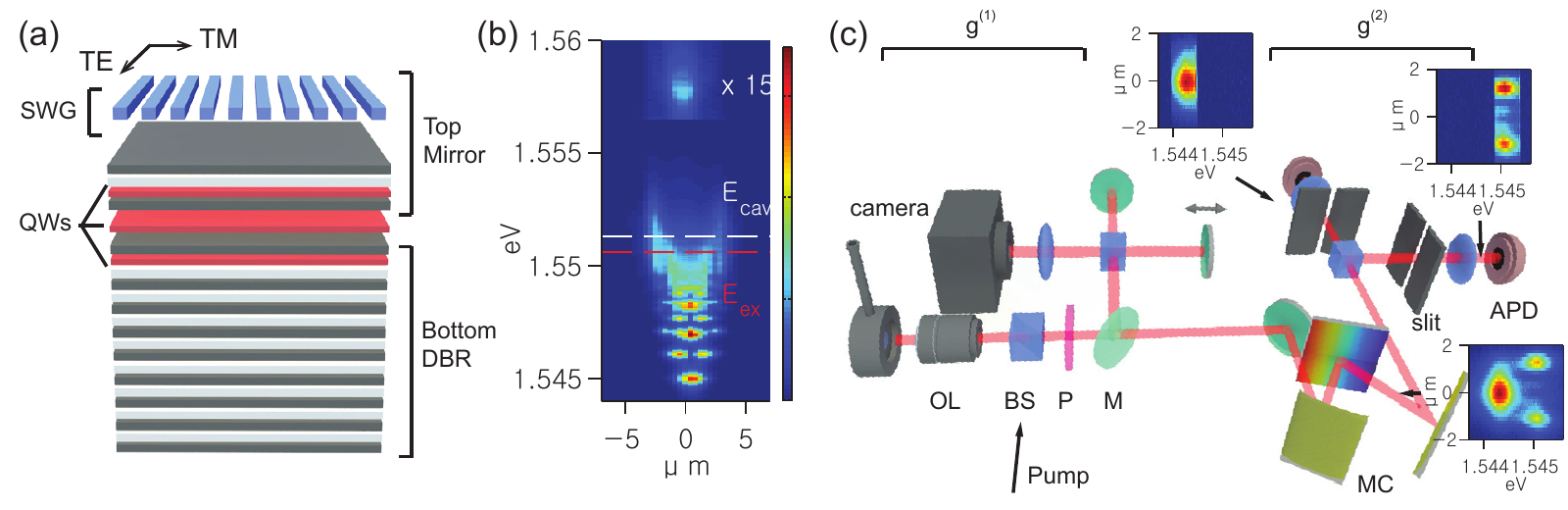}
\caption{The cavity system and experimental setup. (a) A schematic of the SWG-based microcavity.  (b) The spectrally resolved real space image of polariton states at a low excitation power. $\mathrm{E_{cav}}$ is the cavity resonance and $\mathrm{E_{ex}}$ is the exciton resonance. $\mathrm{E_{cav}}$ is estimated by using lower polariton, upper polariton, and exciton energies. (c) A schematic of the experimental setup. OL: Objective lens, BS: Beam splitter, P: Polarizer, M: Mirror, MC: Monochromator, and APD: Avalanche photodetector. For the $g^{(2)}$ measurements, a monochromator followed by two mechanical slits was used to spectrally filter the discrete polariton states. Examples are shown for the spectrally-resolved real space images of the lowest two LP states right after the monochromator (bottom), the spectrally filtered ground state (top left) and the spectrally filtered first excited state (top right). The resolution of the spectral filter was about 0.08~nm, determined by the monochrometer resolution.
}
\label{fig:schematic}
\end{figure*}

The cavity used in our work has both high polarization selectivity and tight lateral confinement, so as to suppress mode-competition and condensate depletion while enhancing condensate nonlinearity. As illustrated in Fig.~1(a), a suspended high index contrast sub-wavelength grating (SWG) is used as the top mirror \cite{Huang_surface-emitting_2007}, in place of a more commonly used distributed Bragg reflector (DBR). The grating breaks the rotational symmetry and was designed to allow high reflectance for only the transverse-electric (TE) polarized mode \cite{schablitsky_controlling_1996, Huang_surface-emitting_2007}. Hence polariton modes were formed in a single, predetermined spin state \cite{zhang_zero-dimensional_2014}. This eliminates mode competition between different spin states that is ubiquitous in DBR-DBR cavities. The high index contrast allows high reflectance even with SWGs of a few wavelengths in size. Hence polariton modes were tightly confined to within the SWG region of 7.5~$\mathrm{\upmu m}$ $\times$ 7.5~$\mathrm{\upmu m}$ \cite{zhang_zero-dimensional_2014, zhang_coupling_2015}, featuring a discrete energy spectrum (Fig.~1(b)). The non-degenerate ground state is separated from the first excited state by about $1~\mathrm{meV}$. The energy gap protects the condensate from quantum depletion and mode competition with the excited states. In addition, the discrete energy levels allow us to unambiguously select and measure each individual state, as illustrated in Fig.~1(c). The tight confinement also enhances the nonlinear polariton-polariton interactions.

\section{\textsc{Spectral Properties of the Polariton Laser}}

\begin{figure*} [tb]
\centering
\includegraphics[width=\textwidth]{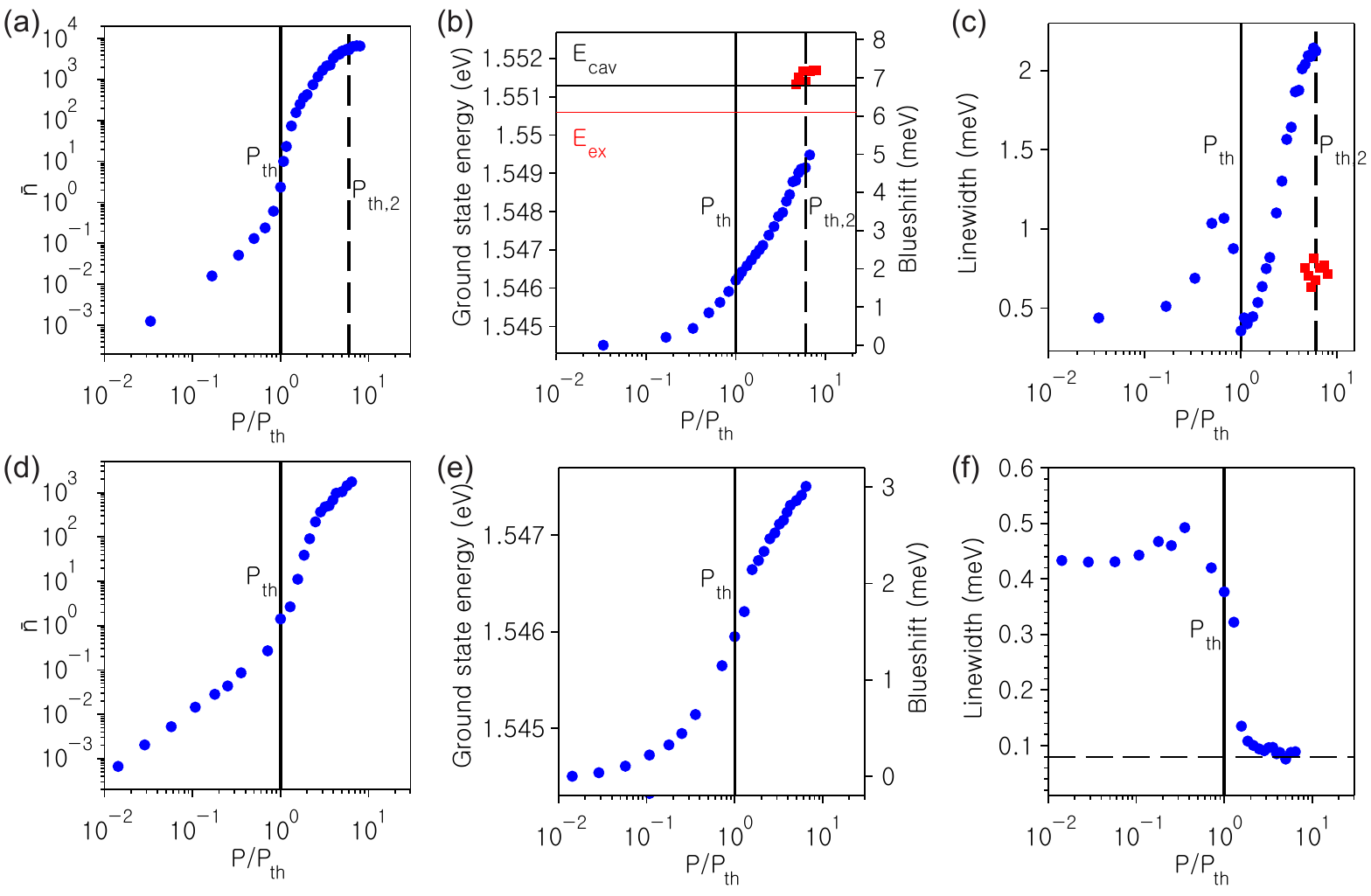}
\caption{Intensity and spectral properties of the single-mode polariton laser. (a) The occupation number vs. $P/P_{th}$ for pulsed excitation. $P_{th}$ indicates the threshold for polariton lasing and $P_{th,2}$ indicates the threshold for photon lasing. $\bar{n}$ is estimated from the independently measured PL intensity from the ground state, collection and detection efficiencies of the setup, and the polariton lifetime \cite{deng_polariton_2003, bajoni_polariton_2008}. (b) Energy and blueshift of the polariton ground state (dots) and lasing photon mode (squares) vs. the normalized pump powers $P/P_{th}$ for pulsed excitation. Here $P_{th}$ is the pump power at the polariton lasing threshold. The solid vertical line marks the polariton lasing threshold and the dashed vertical line marks the photon lasing threshold. (c) The linewidth (full-width at half-maximum) of the polariton ground state vs. $P/P_{th}$ for pulsed excitation. (d), (e), (f) The occupation number, blueshift, and linewidth of the polariton ground state vs. $P/P_{th}$ for CW excitation. The dashed line in (F) represents the spectral resolution of the monochromator of about 0.08~nm.
}
\label{fig:lasing}
\end{figure*}

With increasing excitation density $P$, polariton lasing was observed by a sharp super-linear increase of the ground state population $\bar{n}$ around the threshold $P_{th}$ (Fig.~\ref{fig:lasing}(a) and (d)), corresponding to the onset of quantum degeneracy in the ground state, or $\bar{n}=1$. At the same time, the spectral linewidth of the ground-state polariton emission narrowed, as shown in Fig.~\ref{fig:lasing}(c) and (f), reflecting increased phase coherence. The accurate linewidth and coherence time of the polariton laser were measured by a Michelson interferometer with continuous-wave (CW) excitation, as we will discuss later.

Additional confirmation of polariton lasing, rather than photon lasing, was shown by the emission energy $E_{LP}$ of the polariton laser vs. $P$ in Fig.~\ref{fig:lasing}(b) and (d). The energy of the polariton laser corresponds to the resonance energy of the polariton ground state, hence it blue-shifted with increasing $P$ due to phase space filling and polariton interactions. Moreover, the blueshift was continuous across the polariton lasing threshold and $E_{LP}$ remained well below the cavity or exciton resonances up to many times above the threshold, confirming that the strong-coupling regime was maintained.

In contrast, drastically different spectral properties were observed when a transition to photon lasing in the weak-coupling regime took place at $P\sim 6P_{th}$ under pulsed excitations. The transition was marked by a sudden appearance of a new lasing mode pinned at the energy of the cavity resonance with a sharply decreased linewidth (red symbols in Fig.~\ref{fig:lasing}(b) and (c)).

\section{\textsc{Intensity noise of the Polariton Laser}}

\begin{figure*} [tb]
\centering
\includegraphics{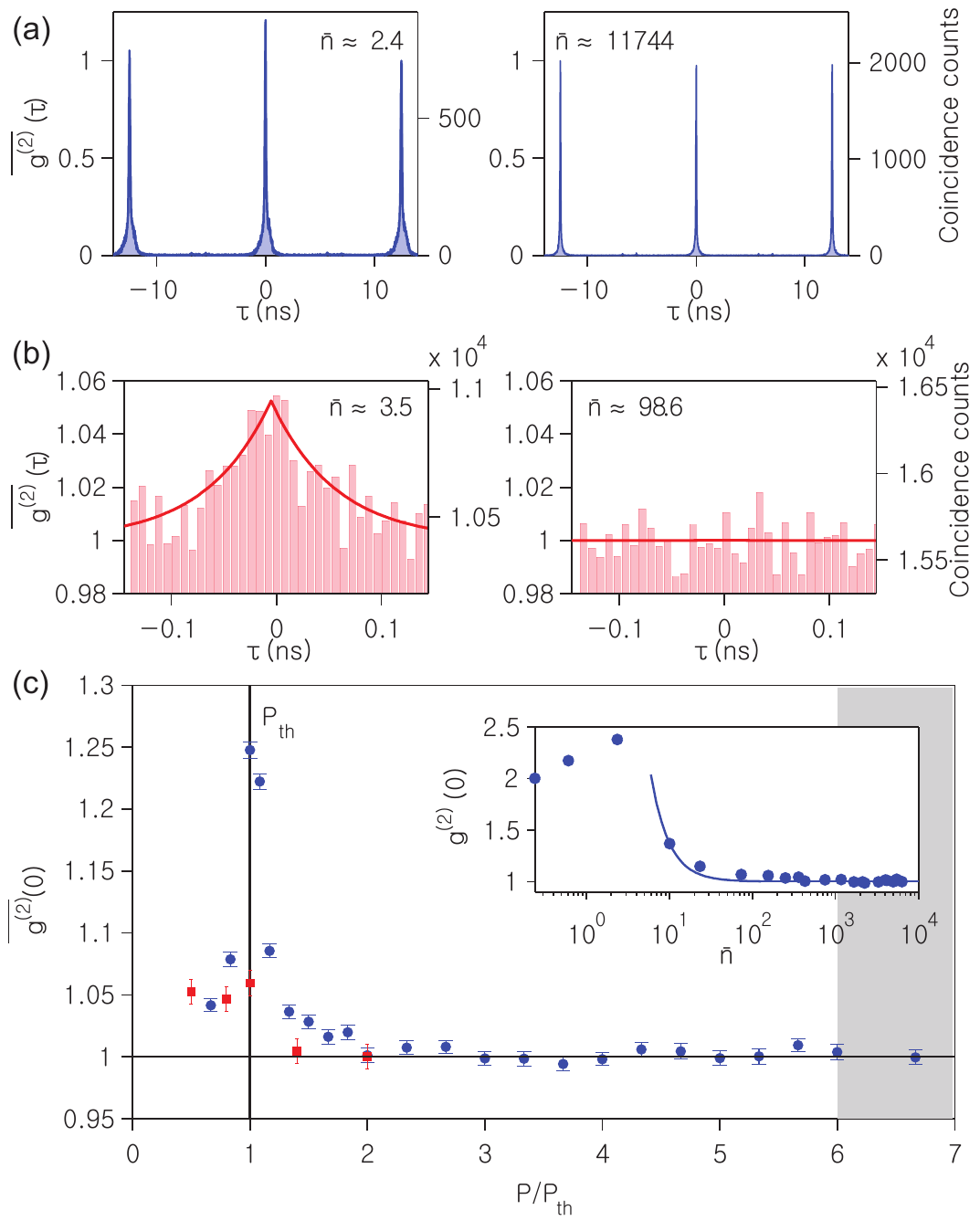}
\caption{Second-order coherence properties of the single-mode polariton laser. (a) $\overline{g^{(2)}}(\tau)$ vs. $\tau$ below and above the threshold for pulsed excitation. (b) $\overline{g^{(2)}}(\tau)$ vs. $\tau$ below and above the threshold for CW excitation. (c) $\overline{g^{(2)}}(0)$ vs. $P/P_{th}$ for pulsed (dots) and CW (rectangles) excitations. The error bars indicate statistical error of one standard deviation. The grey-shaded area shows where the polariton and photon lasing coexist. Inset: $g^{(2)}(0)$ vs. $\bar{n}$ of pulsed excitation corrected for the relaxation time of the ground state \cite{supp}. The solid line shows a theoretical fit by Eq.~(\ref{eq:1}), yielding $n_s = 37.3\pm0.9$.
}
\label{fig:singleg2}
\end{figure*}

We characterized the intensity noise of the polariton laser by measurements of $g^{(2)}(\tau)$ using a Hanbury-Brown Twiss (HBT) type of setup (Fig.~1(c)), under both pulsed and continuous-wave (CW) excitations.

The measured value $\overline{g^{(2)}}(0)$ is an integration of ${g^{(2)}}(\tau)$ over the measurement time $\Delta T$. $\Delta T$ is the duration of the emission pulse for pulsed excitation or the detector time resolution of 40~ps for CW excitation. For a single-mode polariton laser, the ${g^{(2)}}(\tau)$ vs. $\tau$ relation is known. Therefore the actual deviation from the shot-noise limit, $|{g^{(2)}}(0)-1|$, or its upper bound can be obtained accurately from $\overline{g^{(2)}}(0)$ \cite{supp}.

Examples of the measured $\overline{g^{(2)}}(\tau)$ vs. $\tau$ are shown in Fig.~\ref{fig:singleg2}(a) and (b). For both pulsed and CW excitations, bunching was evident below threshold but absent above threshold, showing the transition to a coherent state above threshold. The variation of $\overline{g^{(2)}}(0)$ with the normalized excitation power $P/P_{th}$ was shown in Fig.~\ref{fig:singleg2}(c), with the corresponding $g^{(2)}(0)$ vs. $\bar{n}$ shown in the inset.

A rapid transition from a thermal to coherent state was evident. Near $P_{th}$, where the ground-state occupation number $\bar{n}$ was small, bunching was measured with $\overline{g^{(2)}}(0)$ as high as $1.248\pm0.007$ under pulsed excitations, corresponding to $g^{(2)}(0)\sim 2$ after correcting for the time average \cite{supp}.
With the onset of quantum degeneracy and sharp increase of $\bar{n}$ with $P$, the intensity noise rapidly decreased toward the coherent limit. Between $2 P_{th}$ and $6P_{th}$, with condensate occupation number $\bar{n} = 10^2 - 10^3$, the measured and corrected values of intensity noise remained around unity, with $0.994\pm0.006 \le \overline{g^{(2)}}(0) \le 1.009\pm0.005$ ($0.988\pm0.012 \le g^{(2)}(0) \le 1.020\pm0.011$), and the average intensity noise in this range was $\overline{g^{(2)}}(0) = 1.002 \pm 0.002$ ($g^{(2)}(0) = 1.004 \pm 0.004$).  These results demonstrate the rapid formation of a coherent state with Poisson intensity noise in a polariton laser. 

The experimental data were very well described by an analytical model for single-mode matter-wave lasers \cite{wiseman_defining_1997, thomsen_atom-laser_2002, love_intrinsic_2008, whittaker_coherence_2009}. The model includes the interaction within the lasing mode, or in our system the self interaction among the condensed polaritons. It also includes other essential mechanisms of a laser: gain, gain saturation, and decay of the lasing mode. In the Bose-degenerate limit of $\bar{n} \gg 1$, $g^{(2)}(0)$ can be obtained as: \cite{wiseman_defining_1997,whittaker_coherence_2009}:
\begin{equation}
g^{(2)}(0)=1+\frac{n_s} {\bar{n}^2},
\label{eq:1}
\end{equation}
where $n_s$ is the gain saturation number. 

\begin{figure} [htb]
\centering
\includegraphics{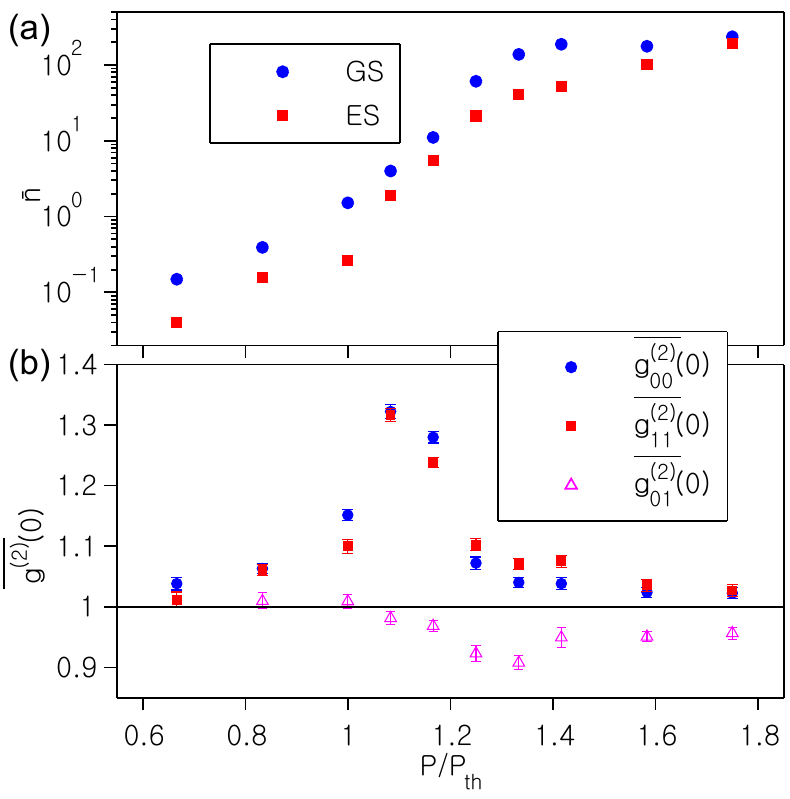}
\caption{Second-order correlations of the two-mode polariton laser. (a) The occupation numbers of the ground state (dots) the excited state (squares) vs. $P/P_{th}$. GS: Ground state, ES: First excited state. (b) $\overline{g^{(2)}}(0)$ vs. $P/P_{th}$. $\overline{g^{(2)}_{00}}(0)$: auto-correlation of GS, $\overline{g^{(2)}_{11}}(0)$: auto-correlation of ES, and $\overline{g^{(2)}_{01}}(0)$: cross-correlation between GS and ES.
}
\label{fig:multig2}
\end{figure}

Comparing Eq.~(\ref{eq:1}) and Eq.~(\ref{eq:g2}) shows that the total number fluctuations in the condensate is $\sigma_n^2 = n_s + \bar{n}$; $n_s$ represents fluctuations induced by the reservoir and other non-condensed modes, while $\bar{n}$ represents the intrinsic shot-noise of the condensate. The coherent limit is reached when $\bar{n} \gg \sqrt{n_s}$. Fitting our data with Eq.~(\ref{eq:1}) gives $n_s = 37.3\pm0.9$, with $g^{(2)}(0)-1< 10^{-3}$ at $P>2.5 P_{th}$ (inset of Fig.~\ref{fig:singleg2}(c)).

Our result is in sharp contrast to previous 2D polariton lasers. Previously, slow decrease of $g^{(2)}(0)$ with $P$ toward a value above unity was commonly observed, suggesting large $n_s$. For example, in a 2D system featuring multiple localized lasing modes with long coherence times, $g^{(2)}(0)=1.1$ was obtained for the selected lasing mode with $\bar{n}\sim 500$. Correspondingly, $n_s = 25,000 \gg \bar{n}$ dominates the intensity noise \cite{love_intrinsic_2008}. In other experiments, $g^{(2)}(0)$ was typically higher or could not be obtained accurately. 

Since a condensate population of $10^2-10^4$ was commonly reported when the transition to the weak-coupling regime takes place, a relatively small $n_s$ as shown here is crucial for establishing intensity stability in a polariton laser.

In addition to reservoir induced intensity fluctuations, we show the effect of mode competition on $g^{(2)}(0)$ in a multi-mode polariton laser. We established two-mode lasing in the same system under the same experimental conditions except for moving the excitation laser spot from the center of the device to slightly off-center, to increase its overlap with the first excited state of the polariton. \cite{bajoni_polariton_2008}. The input-output relationships of both lasing modes are shown in Fig.~\ref{fig:multig2}(a).  Contrary to a single-mode laser, clear deviations of $g^{(2)}(0)$ from the unity were observed for both the ground and first-excited state, with $\overline{g^{(2)}_{00}}(0)$ ($g^{(2)}_{00}(0)$) $= 1.023\pm0.009$ $(1.048\pm0.019) $ and $\overline{g^{(2)}_{11}}(0)$ ($g^{(2)}_{11}(0)$) $= 1.027\pm0.009$ $(1.057\pm0.019)$, respectively (Fig.~\ref{fig:multig2}(b)). The increased intensity fluctuations can be explained by the stochastic relaxation of polaritons from a common reservoir into the lasing modes \cite{kusudo_stochastic_2013}. Consistent with this explanation, we measured strong anti-correlation between the two modes as shown by a cross-correlation function $\overline{g^{(2)}_{12}}(0) < 1$.  We note that such mode competition is difficult to eliminate in 2D or quasi-2D systems, because the linewidth of the lasing mode is typically larger or comparable to the energy separation between LP modes of different polarizations, of different momenta, or in different localization potentials. Our results show that single-mode lasing, achieved by both tight lateral confinement and polarization selectivity in our system, is also crucial for intensity stability of a polariton laser.

\section{\textsc{Phase noise and the condensate interaction energy}}

\begin{figure*} [tb]
\centering
\includegraphics{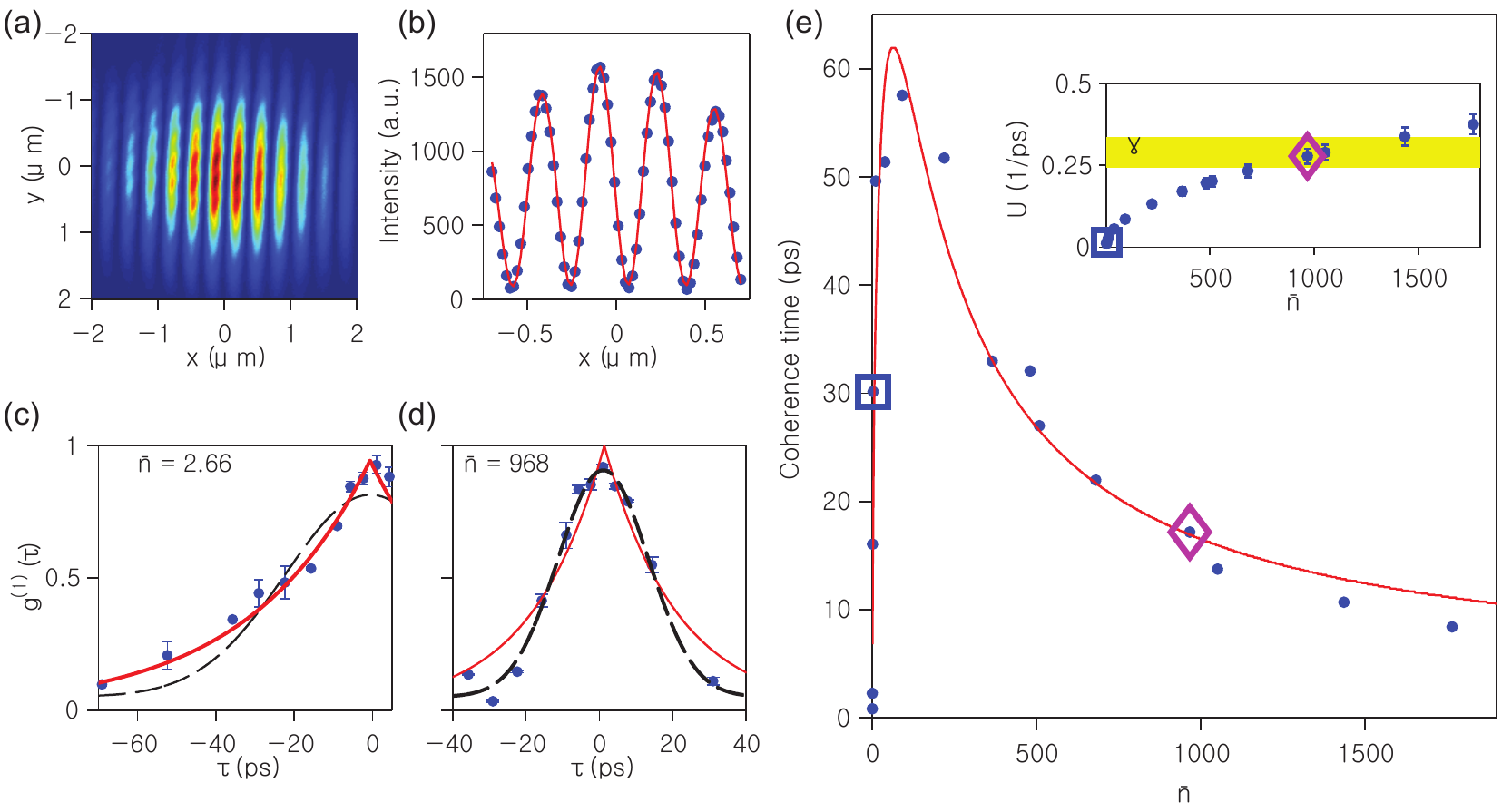}
\caption{First-order coherence properties of the single-mode polariton laser. (a) A spatial interference image of the polariton laser at zero time delay. (b) The interference fringe along x-axis obtained from (a) by integration along the y-axis. The dots are data and the solid line is a fit by equation~(\ref{eq:A1}) with $g^{(1)}(0) = 0.94$ (see Appendix). (c), (d) The measured $g^{(1)}(\tau)$ vs. $\tau$ (dots), exponential fits (red solid lines), and Gaussian fits (black dashed lines), for $\bar{n} = 2.66$ and $968$, respectively. (e) The phase coherence time $\tau_c$ of the polariton laser vs. $\bar{n}$. The dots are taken from the measured $g^{(1)}(\tau)$ for each value of $\bar{n}$. The line is calculated from Eq.~(\ref{eq:2}) using the fitted parameters.  Inset: Comparison between the interaction energy $U$ (dots) and decay rate $\gamma$ (yellow shaded region) of the condensate vs. $\bar{n}$. The error bars and thickness of the shaded area are determined by fitting errors with a 95\% confidence interval. The rectangle and diamond marks respresent $\bar{n} = 2.66$ and $\bar{n} = 968$, respectively.
}
\label{fig:singleg1}
\end{figure*}

The temporal phase coherence of a polariton laser is described by the first-order correlation function, $g^{(1)}(\tau)$.  It is related to the power spectrum of the polariton emission by a Fourier transform.
Using an intensity-stabilized CW laser for excitation, we measured $g^{(1)}(\tau)$ of the single-mode polariton laser using a Michelson interferometer (Fig.~1(c)). The visibility of the interference fringes gives $g^{(1)}(\tau)$ (see Appendix). Examples of the interference fringes are shown in Fig.~\ref{fig:singleg1}(a) and (b). We then obtain $g^{(1)}(\tau)$ vs. $\tau$ at different excitation powers by varying the delay time $\tau$ between the two arms of the interferometer.

Very different $\tau$ dependence of $g^{(1)}(\tau)$ was observed for the polariton laser as the condensate occupancy was increased. As show in Fig.~\ref{fig:singleg1}(c), just above threshold, at a low condensate occupancy of $\bar{n} \sim 2.7$, $g^{(1)}(\tau)$ decays exponentially with $\tau$ (Fig.~\ref{fig:singleg1}(c)), corresponding to a Lorentzian line shape. This confirms single-mode lasing and shows clearly that the intrinsic dephasing of the condensate dominates over external effects. 

As $\bar{n}$ increases, $g^{(1)}(\tau)$ changed to Gaussian decay, as is apparent in Fig.~\ref{fig:singleg1}(d) for $\bar{n} = 968$. The coherence time also decreased. This can not be explained by multi-mode lasing or extrinsic effects, which were excluded by the exponential decay at small $\bar{n}$. It is also distinct from photon lasers, where exponential decay of $g^{(1)}(\tau)$ and Lorentzian spectrum persists in single-mode lasers.

The transition to a Gaussian decay of $g^{(1)}(\tau)$ can be described using the same single-mode matter-wave laser theory as used to describe the $g^{(2)}$ measurements. The $g^{(1)}(\tau)$ of a matter-wave laser is given by \cite{ thomsen_atom-laser_2002, whittaker_coherence_2009}:
\begin{align}
\label{eq:2}
|g^{(1)}(\tau)| =& \exp{\Big(-\frac{4(n_s + \bar{n})u^2}{{\bar{\gamma}}^2}\big(\exp{(-\bar{\gamma}\tau)}+\bar{\gamma}\tau-1\big)\Big)}\nonumber \\
&\times \exp{\Big(\frac{n_s + \bar{n}}{4{\bar{n}}^2}\big(\exp{(-\bar{\gamma}\tau)}-\bar{\gamma}\tau-1\big)\Big)}, \\
\bar{\gamma} =& \frac{\bar{n}}{n_s + \bar{n}}\gamma. \nonumber
\end{align}
Here $u$ is the polariton-polariton interaction constant and $\gamma$ is the decay rate of the polariton ground state. This equation can be simplified in two limits corresponding to the weak and strong interaction regimes. To separate the two regimes, we define a total interaction strength $U = 4u\sqrt{\bar{n}}$. In the weak-interaction regime, $U \ll \gamma$, equation~(\ref{eq:2}) is reduced to the Schawlow--Townes formula for a photon laser, $\exp{(-\gamma\tau/{2\bar{n}})}$, featuring an exponential decay (Fig.~\ref{fig:singleg1}(c)). Correspondingly, the $1/e$ coherence time increases linearly with the occupation number of the lasing mode, $\tau_c = 2\bar{n}/\gamma$. In the strong-interaction regime, $U \gg \gamma$, equation~(\ref{eq:2}) is approximated by $\exp{(-2\bar{n}u^2\tau^2)}$, featuring a Gaussian decay, as we observed (Fig.~\ref{fig:singleg1}(d)). Correspondingly, $\tau_c\propto 1/U\propto 1/\sqrt{\bar{n}}$. Therefore polariton-polariton interactions and the shot noise of the condensate lead to a Gaussian broadening or dephasing of the polariton laser.

Using Eq.~(\ref{eq:2}) to fit the $g^{(1)}(\tau)$ data at different $\bar{n}$, we can obtain the main parameters governing the dynamics of the polariton laser, $n_s$, $\gamma$ and $u$. The best fit yields $n_s = 61\pm13$, $\gamma= 0.29\pm0.04$~$\mathrm{ps^{-1}}$ and $u = (2.2\pm0.2) \times 10^{-3}$~$\mathrm{ps^{-1}}$. $n_s$ is similar to the estimate obtained from pulsed $g^{(2)}$ measurements, 37.3. The difference may be due to different excitation conditions, CW vs. pulsed, which may lead to different densities and effective temperatures of the reservoir. $\gamma$ is within a reasonable range. From $u$ we estimated the system size-independent interaction constant $uA_{\mathrm{cond}}=4$~$\mathrm{\upmu eV}\cdot\mathrm{\upmu m^2}$, which is in excellent agreement with the previously reported theoretical and experimental values \cite{ferrier_interactions_2011}. Here $A_{\mathrm{cond}}\sim2.5~\mu$m$^2$ is the size of the condensate measured from spatial PL imaging (see the inset of Fig.~\ref{fig:schematic} for an example). It was independent of the pump power as expected, since it was determined by the effective confinement potential in 0D systems. This confirms that strong polariton-polariton interaction $u$ could be achieved in our system due to the tight lateral confinement or small $A_{\mathrm{cond}}$.

Figure~\ref{fig:singleg1}(e) compares $\tau_c$ from the fit with experimental values with respect to $\bar{n}$. For small $\bar{n}$, $\tau_c$ increases sharply with $\bar{n}$, as expected from the Schawlow-Townes formula. However, $\tau_c$ starts to decrease because polariton-polariton scattering leads to the phase decoherence of the lasing mode. The crossover between the weak and strong interaction regimes corresponds to where $U\sim \gamma$, as illustrated in the inset of Fig.~\ref{fig:singleg1}(e).

We note that, although linewidth broadening has been observed in polariton lasers before, contribution by the condensate nonlinearity was negligible. Typically the linewidth broadening was accompanied by an increase in the intensity noise \cite{kasprzak_second-order_2008,  tempel_characterization_2012}, and thus could be understood as the effect of mode competition. When a single, localized mode was selected, the coherence time saturated above threshold and became independent of the ground state occupancy \cite{love_intrinsic_2008}; the coherence time was mainly limited by energy shift resulting from reservoir-induced thermal fluctuations represented by $n_s\gg \bar{n}$. Here, however, the intensity noise remained at the shot-noise limit and thus multi-mode lasing or reservoir induced fluctuations were both negligible. $g^{(1)}(\tau)$ showed strong dependence on the condensate population. Therefore, the Gaussian dephasing we observed directly resulted from interactions within the condensed polaritons.

We also note that dephasing in the condensate may also be induced by thermal fluctuations of the reservoir population \cite{love_intrinsic_2008, kavokin_microcavities_2007}. However, this effect would not explain the fast decrease of the coherence time above 2$P_{th}$, and thus was expected to be much weaker compared to the condensate’s contribution.  As shown in Fig.~2(e), the energy shift vs. excitation density, $d\Delta E/dP$, between $2-12P_{th}$ was slowed down to 1/10 of that below threshold. Therefore the energy fluctuation due to reservoir population fluctuation would change by $\ll \sqrt{6}$ between $2-12 P_{th}$, which is in direct contradiction to the observed 6-fold decrease of the coherence time in this range.

The reason the reservoir-induced fluctuation was weak could be many-fold.  First of all, while the energy shift is proportional to the population $N$, the energy fluctuation is proportional to $\sqrt{N}/A$, where A is the system area.  In our system, there was no lateral confinement in the QWs and the excitons can freely diffuse. A typical diffusion length \cite{tosi_sculpting_2012} gives a spatial extend of about 100~$\mathrm{\upmu m^2}$. In contrast, the polaritons were tightly confined; their spatial extend was determined by the ground-state wavefunction and measured to be 2.5 $\mathrm{\upmu m^2}$ for all excitation densities.  Hence, the energy fluctuation introduced by an unconfined exciton population is attenuated by 1600 times compared to a confined one.  At an exciton density of 0.1$n_{sat}$, for a saturation density $n_{sat}=4\times10^{10}~\mathrm{cm^{-2}}$ per QW \cite{Houdre_saturation_1995}, the exciton has a total population of about $4.8\times10^4$, which introduces an energy fluctuation equivalent to that by about 30 condensate polaritons near zero detuning, while the condensate population quickly built up to $10^2-10^3$ above threshold. Moreover, it would be interesting to investigate if the coherent condensate interacts within itself more strongly than with the thermal reservoir, and if the reservoir fluctuation induced dephasing may become suppressed when the coherent condensate is formed \cite{Haug_coherence_2010}. These issues could be clarified in future investigations with more careful calibration of the exciton density or exciton interaction strength.

\section{\textsc{Conclusion}}
In conclusion, we have demonstrated the first polariton laser with shot-noise limited intensity stability, or full second-order coherence, operating as a single-mode laser in the gain saturation regime. At high lasing intensities, a transition from exponential to Gaussian decay of its intrinsic temporal phase coherence was observed, which can be understood as resulting from strong interactions within the lasing mode, the polariton condensate.
Experimental results of the phase and intensity noise were well described by an analytical model of single-mode matter-wave lasers, which yielded the basic parameters governing a polariton laser's dynamics, including a small gain saturation number and a large polariton interaction strength. The demonstrated intensity stability is a critical feature for lasers. The interaction-induced change in $g^{(1)}(\tau)$ unambiguously reveals the matter-wave origin of the polariton laser. The strong polariton interactions will be important for nonlinear polariton devices \cite{verger_polariton_2006,liew_optical_2008,liew_single_2010}.

\section{\textsc{APPENDIX: Materials and Methods}}
\setcounter{equation}{0}
\renewcommand{\theequation}{A\arabic{equation}}

The sample consists of 3 stacks of 4 GaAs/AlAs QWs placed at the central three anti-nodes of a $\lambda/2$, AlAs cavity. The bottom mirror of the cavity is formed by a DBR consisting of 30 pairs of Al$_{0.15}$Ga$_{0.85}$As/AlAs layers. The top mirror of the cavity if formed by an Al$_{0.15}$Ga$_{0.85}$As SWG suspended over a three-layer top DBR. The planar wafer was grown by molecular beam epitaxy. The SWG was created by patterning via electron-beam lithography followed by a reactive ion etching. Then a selective wet etching process followed by a critical point drying was used to remove an Al$_{0.85}$Ga$_{0.15}$As sacrificial layer to suspend the grating. We directly measured the energies of the weakly coupled TM excitons and the lower and upper polaritons from their PL, as shown in (Fig.~\ref{fig:schematic}(b)), which gave a Rabi splitting of 12~meV and detuning of about 0.7~meV. The estimated quality factor of the sample from the linewidth measured at low pump power was about 4,000.

The sample was kept at 10~K and excited by a pulsed or continuous-wave (CW) Ti:S laser. The pulsed laser had a pulse width of 150-fs and a repetition rate of 80~MHz. The CW laser was frequency-locked to within 100~KHz bandwidth and chopped by an electro-optic modulator with 10\% duty cycle at 1~MHz to reduce sample heating. The energy of the pump laser was tuned at least 20~meV above the exciton resonance to avoid any coherence induced by the pump laser. $g^{(1)}$ experiments were done using the CW laser and $g^{(2)}$ experiments were done using both the pulsed and CW lasers. An objective lens was used to focus the pump laser to a spot 2~$\mathrm{\upmu m}$ in diameter and collect the PL from the sample. The emission was then either sent to a Michelson interferometer or Hanbury Brown and Twiss (HBT) setup for the $g^{(1)}$ and $g^{(2)}$ measurements, respectively.

For $g^{(1)}$ measurements, The intensity distribution in Fig.~\ref{fig:singleg1}(b) can be described by
\begin{align}
\label{eq:A1}
I_{CCD}(x,\tau) = &I_{1}(x) + I_{2}(x) + 2|g^{(1)}(\tau)| \nonumber \\
& \times \sqrt{I_{1}(x)I_{2}(x)}cos(\frac{2\pi \theta}{\lambda_{0}}x + \phi),
\end{align}
where $\lambda_0$ is the wavelength of the lasing mode, $\phi$, $\theta$ and $\tau$ are the phase difference, angle and time delay between the two interfering beams at given $\tau$. $I_{1}(x)$ and $I_{2}(x)$ are the Gaussian intensity profile of the two beams, respectively, and are equal with $<1\%$ difference in amplitude. Fitting the measured interference patterns with Eq.~(\ref{eq:A1}), we obtained $g^{(1)}(\tau)$ for each $\tau$. Varying the excitation power, we obtained the power dependence of $g^{(1)}(\tau)$ vs. $\tau$. $g^{(1)}(0)\ge 0.9$ was maintained throughout the experiments.

\section{Acknowledgments}
We thank Prof.~Paul Eastham for helpful discussions. SK, ZB, ZW, and HD acknowledge the support by the National Science Foundation (NSF) under Awards DMR 1150593 and the Air Force Office of Scientific Research under Awards FA9550-15-1-0240. CS, SB, MK and SH acknowledge the support by the State of Bavaria, Germany. The fabrication of the SWG was performed in the Lurie Nanofabrication Facility (LNF) at Michigan, which is part of the NSF NNIN network.

\newpage
\setcounter{equation}{0}
\renewcommand{\theequation}{S\arabic{equation}}
\subsection*{Supplemantry Information for ``A Coherent Polariton Laser"}
\noindent\textbf{\textrm{Estimating $g^{(2)}(0)$ and its upper and lower bounds}}.\hspace{10pt}  The measured auto-correlation function, $\overline{g^{(2)}}(0)$, is an average of the actual $g^{(2)}(\tau)$ over the time resolution of the measurement. Hence when the time resolution is much longer than the intensity correlation time, the measured $\overline{g^{(2)}}(0)$ approaches $1$ due to averaging. When the time resolution is shorter than the intensity correlation time, $\overline{g^{(2)}}(0)$ approaches the actual $g^{(2)}(0)$. In CW measurements, the time resolution is determined by the response time of the photon counters, which was measured to be $\sim$40~ps for both of our counters. In pulsed measurements, often $\overline{g^{(2)}}(0)$ is obtained by integration over the whole pulse and thus the time resolution is determined by the duration of the measured pulses, $\Delta T$. In our experiments, the polariton emission pulse shortened rapidly from $\gg 40$~ps below threshold to $<$4~ps above threshold due to the stimulated scattering (Fig.~6(a)). Correspondingly, an increase of $\overline{g^{(2)}}(0)$ near the threshold was observed (Fig.~2(c) in the main text).

\begin{figure} [b]
\centering
\includegraphics[width=\textwidth]{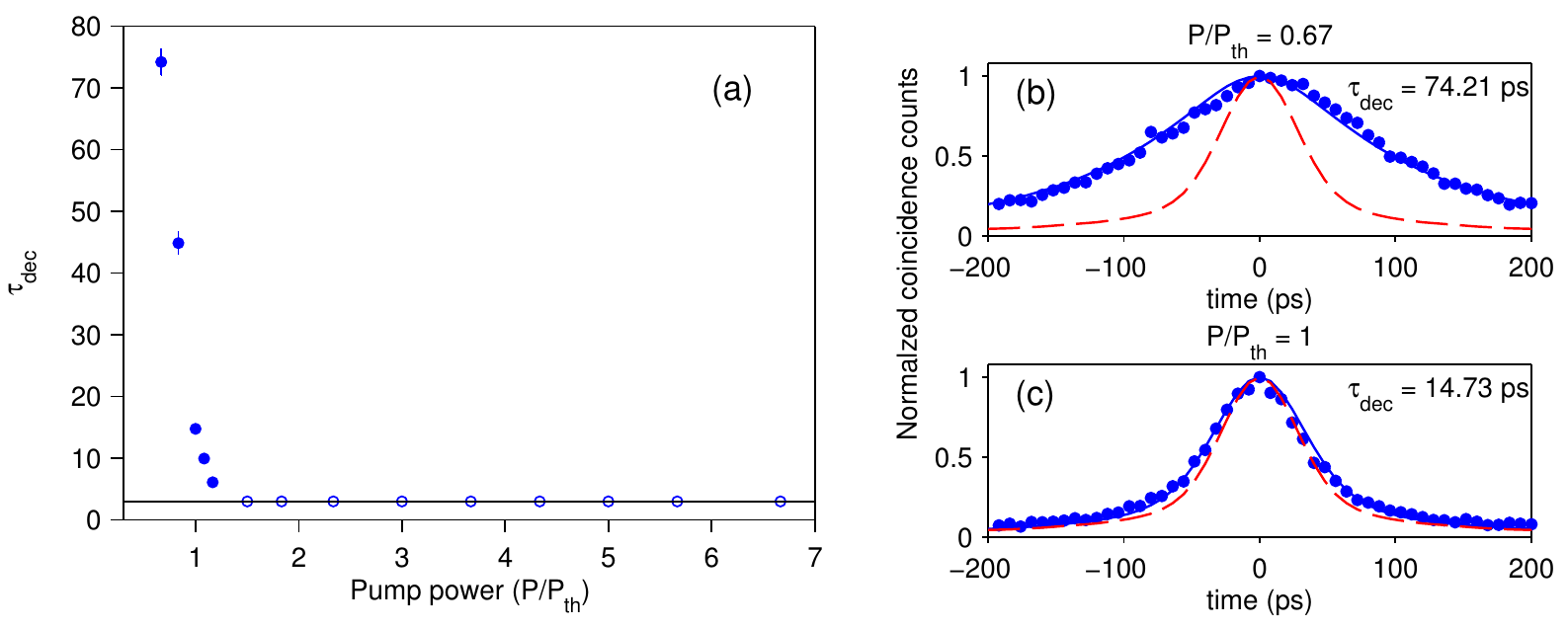}
\caption{(A) The pulse duration of the ground state polariton emission $\Delta T$ vs. the normalized pump power $P/P_{th}$. $\Delta T$ are obtained by fitting $g^{(2)}(\tau)$ measured for two uncorrelated pulses, taking into account the measured IRFs of the photon counters. $\Delta T$ was well fitted at $P/P_{th} < 1.5$ (dots) and unresolvable at $P/P_{th} > 1.5$ (circles). The solid line is the intensity correlation time of 3.1~ps estimated from $g^{(2)}(0)$ below the threshold. (B), (C) Two examples of the least square fitting of $g^{(2)}(\tau)$ vs. $\tau$ of two uncorrelated pulses, at $P/P_{th} = 0.67$ and 1, respectively. The dots are data; the solid lines are the fits using the measured IRF; the dashed lines are the convolution of two IRFs of the photon counter.
}\label{fig:S1} 
\end{figure}

To evaluate the actual $g^{(2)}(0)$ from $\overline{g^{(2)}}(0)$, we need to know the functional form of $g^{(2)}(\tau)$ and the time resolution. Note that, for pulsed experiment, since $\overline{g^{(2)}}(0)$ depends on the convolution of the pulse, it is also largely determined by time resolution, or the pulse duration $\Delta T$, and is insensitive to the actual shape of the pulse. Below we will approximate the emission pulse by an exponentially decaying pulse. In this way, a single parameter, the 1/e decay time, capture the time duration $\Delta T$, and an analytical relation between $g^{(2)}(0)$ and $\overline{g^{(2)}}(0)$ can be obtained. We have also checked that assuming a Gaussian (rather than exponential) pulse shape would alter our estimates of $\Delta T$ and $|g^{(2)}(0)-1|$ by only a few percents below or near threshold and no difference for $P>1.16P_{th}$.

For a single-mode polariton lasers, the functional form of $g^{(2)}(\tau)$ is the same as that of a standard laser \cite{scully_quantum_1997, whittaker_coherence_2009}:
\begin{equation}
g^{(2)}(\tau) = 1 + \frac{n_s}{\bar{n}^2}\exp{(-\frac{\bar{n}}{n_s+\bar{n}}\gamma \tau)}. 
\nonumber
\label{eq:S1}
\end{equation}
For an exponential pulse with $1/e$ decay time of $\Delta T$, we have \cite{loudon_quantum_2000}:
\begin{align}
\overline{g^{(2)}}(0) &= 1 + (g^{(2)}(0)-1)\frac{1}{\Delta T^2} \int_{0}^{\infty}{\int_{0}^{\infty}{\exp{(\frac{x+y}{\Delta T})}\exp{(\frac{-|x-y|}{\Delta T})}}dxdy } \nonumber \\
& = 1+ (g^{(2)}(0)-1)(\frac{\tau_{c}}{\Delta T+\tau_c}).
\label{eq:S2}
\end{align}
Here $\tau_c = 1/\gamma$ below threshold and $\tau_c=(n_s+\bar{n})/(\bar{n}\gamma)$ above threshold. Hence for our single-mode polariton laser, $g^{(2)}(0)$ can be obtained from $\overline{g^{(2)}}(0)$ given $\Delta T$, $\gamma$ and $n_s$.

To obtain $\Delta T$, we note that:
\begin{align}
  \overline{g^{(2)}}(\tau+nT)\propto \int_{-T/2}^{T/2} {I(t)I(\tau+t+nT)} dt,
\label{eq:S3}
\end{align}
where $n$ is an integer, $T$ is the laser repetition period and $I(t)$ is the convolution of the emission pulse with the instrument response function, IRF$(t)$: $I(t) = \int_{-\infty}^{\infty} {\exp{(-t_{1}/\Delta T)}\mathrm{IRF}(t-t_{1})} dt_{1}$.
We measured the IRF using a pulsed Ti:S laser with a 100~fs pulse duration. Then $\Delta T$ was obtained by a least-square fit of the data with Eq.~\ref{eq:S3} for $\Delta T>4$~ps. The results are shown in Fig.~6(a) and examples of the fit are shown in Fig.~6(b) and (c). At $P > 1.16P_{th}$, $\Delta T <4$~ps, and $\overline{g^{(2)}}(\tau)$ becomes indistinguishable from the convolution of the IRFs. In this regime, we take the conservative limit of $\Delta T=4.2$~ps. It gives the upper bound of $|g^{(2)}(0)-1|$, the deviation from the shot-noise limit.

To obtain $\gamma$, we use the known value of $g^{(2)}(0)=2$ at well below threshold and compare it with $\overline{g^{(2)}}(0)$. Assuming $g^{(2)}(0)=2$ at $P = 0.67P_{th}$ and using $\Delta T = 74.21$~ps (Fig.~6(b)), we find $\gamma=0.31$~ps$^{-1}$  using Eq.~\ref{eq:S2}. This value is consistent with $\gamma=0.29\pm 0.04$~ps$^{-1}$ from the $g^{(1)}$ fitting as discussed in the main text.

Using Eq.~\ref{eq:S2} with $\tau_{c} = 3.23$~ps and $\Delta T$ obtained above, we obtain the corrected $g^{(2)}(0)$ values as shown in the inset of Fig.~3(c) in the main text. At $P > 1.16 P_{th}$, with $\tau_{c} = 3.23$~ps and the conservative estimate of $\Delta T \le 4.2$~ps, we obtain the upper bound of $g^{(2)}(0)- 1\leq 2 (\overline{g^{(2)}}(0)-1) $. Then, for $P=2P_{th}-6~P_{th}$, the maximum (minimum) measured $\overline{g^{(2)}}(0)$ corresponds to corrected $g^{(2)}(0)$ of $1.020\pm 0.011$ ($0.988\pm 0.012$), and the average $\overline{g^{(2)}}(0)$ corresponds to an average $g^{(2)}(0) = 1.004 \pm 0.004$.
%
%
%
%
%
%

\end{document}